\documentclass[12pt]{article}
\usepackage[xdvi]{graphicx}
\newcommand{\bea}{\begin{eqnarray}}
\newcommand{\eea}{\end{eqnarray}}
\newcommand{\Ls}{\left(}
\newcommand{\Rs}{\right)}

\newcommand{\Ll}{\left[}
\newcommand{\Rl}{\right]}
\newcommand{\LL}{\left.}
\newcommand{\RR}{\right.}
\newcommand{\sla}{/\!\!\!}
\newcommand{\nn}{\nonumber}
\newcommand{\abs}[1]{\left| #1 \right|}

\def\del{\partial}
\def\diag{\mathop{\rm diag}\nolimits}

\begin{document}
\setlength{\baselineskip}{0.7cm}
\begin{titlepage}
\begin{flushright}
KOBE-TH-07-05 
\end{flushright}

\vspace{1.0cm}
\begin{center}
{\Large\bf Finite Anomalous Magnetic Moment \\
\vspace*{5mm}
in the Gauge-Higgs Unification} 
\end{center}
\vspace{25mm}

\begin{center}
{\large
Yuki Adachi\footnote{e-mail : yuki1983@kobe-u.ac.jp},
C. S. Lim\footnote{e-mail : lim@kobe-u.ac.jp}
and Nobuhito Maru\footnote{e-mail : 
maru@people.kobe-u.ac.jp}}
\end{center}
\vspace{1cm}
\centerline{{\it Department of Physics, Kobe University,
Kobe 657-8501, Japan}}
%
%   Abstract
%
\vspace{2cm}
\centerline{\large\bf Abstract}
\vspace{0.5cm}
We show that the anomalous magnetic moment of fermion 
in the gauge-Higgs unification 
is finite in {\it any} spacetime dimensions, 
which is a new predictive physical observable similar to the Higgs mass. 
\end{titlepage}

The gauge-Higgs unification \cite{MF, Hosotani} 
is one of the fascinating scenarios 
since it predicts some finite physical observables 
due to the higher dimensional gauge invariance, 
though it is regarded as a nonrenormalizable theory. 
As far as we know, the Higgs mass is the unique finite physical observable 
and its finiteness has been examined from the various points of view 
\cite{HIL}-\cite{Hosotani2}. 
In this scenario, the Higgs scalar field is identified with 
the extra spatial components of the higher dimensional gauge field, 
which immediately forbids the local mass term relying on the gauge invariance. 
Then, the finite Higgs mass is generated by Wilson loop dynamics and 
is independent of the cutoff scale of the theory, 
which is therefore very predictive.

It is a natural question to ask whether 
there are any other finite physical observables 
in the gauge-Higgs unification.  
If we have such an observable, 
we can guess it is in the gauge-Higgs sector of the theory, 
just as the case of the Higgs mass. 
Along this line of thought, 
the divergence structure of S and T parameters has been investigated 
in \cite{LM}, but the local gauge invariant operator for S and T parameters 
is allowed and they are found to be divergent in more than five dimensions 
(although a particular linear combination of them becomes finite 
even in six dimensions).

In this Letter, we find a new finite physical observable 
in {\em any} spacetime dimensions 
in the context of the gauge-Higgs unification. 
It is the anomalous magnetic moment of fermion. 
Before calculating it in detail, 
it is instructive to give an argument of the operator analysis. 
In four dimensions, the gauge invariant dimension six operator 
relevant for the anomalous magnetic moment is given by
\bea
\bar{\Psi}_L \sigma_{\mu\nu} \Psi_R F^{\mu\nu}\langle H \rangle 
+ h.c.
\label{4Dop}
\eea
where $\Psi_{L(R)}$ denotes the standard model fermions 
and $H$ is the standard model Higgs doublet. 
The field strength of photon is denoted by $F^{\mu\nu}$. 
In the gauge-Higgs unification on $M^D \times S^1$ 
where $M^D$ is the $D$ dimensional Minkowski spacetime, 
the Higgs is identified with the extra component of the gauge field $A_y$. 
From the lesson in the discussion for S and T parameters 
in the gauge-Higgs unification \cite{LM}, 
we learn that $A_y$ should be replaced by the extra space component of 
the covariant derivative, $D_y$ in order to preserve 
the higher dimensional gauge invariance. 
This observation leads us to the statement that 
the local gauge invariant operator relevant for the anomalous magnetic moment 
takes the form
\bea
i\bar{\Psi} \sigma_{MN} D_A \Gamma^A \Psi F^{MN}. 
\label{hdop}
\eea
To be precise, the operator describing the effective 3-point vertex 
reads as $i\bar{\Psi} \sigma_{MN} \langle D_A \rangle 
\Gamma^A \Psi F^{MN}$, where $\langle D_A \rangle$ is obtained 
by replacing $A_M$ with its VEV, $\langle A_M \rangle 
= \delta_M^y \langle A_y \rangle$. 
On the other hand, the on-shell condition for the fermion implies 
$\Gamma^M \langle D_M \rangle \Psi = 0$ 
for the fermion without a bulk mass, 
which immediately tells us that 
the local operator describing the magnetic moment vanishes.

Our argument for the finiteness of the anomalous magnetic moment 
does not hold true 
provided the fermion has a gauge invariant bulk mass, 
since the chirality flip is realized by the insertion of the bulk mass, 
even if there is no insertion of the Higgs doublet. 
Though such gauge invariant bulk mass is responsible for QED, 
in a realistic gauge-Higgs unification model incorporating the standard model, 
the zero mode fermions (quarks and leptons) get their masses 
only through the VEV of the Higgs field. 
Namely, our results obtained in this letter remain correct 
in a realistic gauge-Higgs unification.

As for the brane localized operators, 
we notice that the operator 
\bea
i\bar{\Psi} \sigma_{MN} A_y \Gamma^y \Psi F^{MN}
\label{branelocalized}
\eea
is allowed on the branes. 
One might worry that the divergences localized on the branes will appear. 
However, it is known that the shift symmetry 
$A_y \to A_y + {\rm const}$ \cite{GIQ}, 
which is a remnant of the higher dimensional gauge symmetry, is operative 
even at branes. 
Therefore, the brane localized operator (\ref{branelocalized}) 
is forbidden by this shift symmetry.

If this is true, this is a very remarkable result 
since there is a possibility that the anomalous magnetic moment 
is finite in {\em any} spacetime dimensions similar to the Higgs mass, 
in spite of the fact that the higher dimensional gauge theories are argued 
to be nonrenormalizable. 
As far as we know, the finite physical observable other than 
the Higgs mass in the gauge-Higgs unification is not known.

Let us check the above expectation 
by calculating the 1-loop diagrams contributing to 
the anomalous magnetic moment in $(D+1)$ dimensional QED compactified 
on $S^1$ with the radius $R$. 
The action we consider is given by 
\bea
 S = \int d^D x d y \Ll-\frac14 F_{MN}F^{MN} +\bar\Psi i\sla\! D_{D+1} \Psi  
                          +{\cal L}_{GF} \Rl,
\eea
 where $\sla\! D_{D+1} = \sla\! D-i\Gamma_y D_y$, $\Gamma_y^2=1$, 
 $D_M = \del_M - igA_M (M = 0,1,2,3, \cdots ,D)$ is the covariant derivative. 
 $g$ is the $(D+1)$ dimensional gauge coupling constant. 
 The coordinates of $D$ dimensional spacetime and the circle are denoted 
 as $x^\mu~(\mu=0,1,2,3,\cdots,D-1)$ and $y$. 
 We take the metric as $\eta_{MN}=\diag(+,-, \cdots ,-)$.
We choose the gauge fixing term as 
\bea
 {\cal L}_{GF} = -\frac1{2\xi}\Ls \del_\mu A^\mu + \xi \del_y A^y \Rs^2,
\label{GaugeFixing}
\eea
where $\xi$ is a gauge parameter. 
Then, the gauge part of the action becomes 
\bea
 S_G = \int d^D x dy 
       \frac{1}2\Ll -\Ls\del_\mu A_\nu\Rs^2 
             + \Ls1-\xi^{-1}\Rs \Ls\del_\nu A_\nu\Rs^2 
             - \Ls\del_y A_\nu\Rs^2 \RR 
%    \phantom{\int d^4 x dy\frac{1}2} 
       \LL -\Ls\del_\mu A_y \Rs^2 - \xi \Ls\del_y A_y \Rs^2 \Rl.
\eea
Expanding the gauge field in terms of the Kaluza-Klein (KK) modes, 
\bea
 A_M(x^\mu,y) &=& \frac1{\sqrt{2\pi R}} \sum_{n=-\infty}^\infty 
                    A_M^{(n)}(x^\mu) \exp \left(i n\frac{y}{R} \right), 
% A_5(x^\mu,y) &=& \frac1{\sqrt{2\pi R}} \sum_{n=-\infty}^\infty 
%                    A_5^{(n)}(x^\mu) \exp \left(i n\frac{y}{R} \right), 
\eea
 where ${A_M^{(n)}}^*=A_M^{(-n)}$ and integrating out $y$ coordinate, 
 the $D$-dimensional action is written as
\bea
 S_G &=& \int d^D x \sum_{n=-\infty}^\infty \frac{1}{2}
           \Ll -\abs{\del_\mu A_\nu^{(n)}}^2 
             + \Ls1-\xi^{-1}\Rs \abs{\del_\nu A_\nu^{(n)}}^2 
             + {M_n}^2\abs{A_\nu^{(n)}}^2 \RR \nn\\
    %\phantom{\int d^D x \sum_{n=0}^\infty\frac{-1}{2^{\delta_{0n}}}} 
       &&\LL +\abs{\del_\mu A_y^{(n)}}^2 - \xi {M_n}^2\abs{A_y^{(n)}}^2 \Rl,   
\label{gauge}
\eea
 where $M_n= n/R$ is the KK mass. 
The order parameter $\alpha$ is defined as $\langle A_y \rangle \equiv 
-\alpha/(gR)$.\footnote{This sign definition is just a convention. 
This definition is useful because it is easy to check 
whether our results reproduce those of the standard model.} 
Hereafter, our calculation is done in the 't Hooft-Feynman gauge ($\xi=1$).

Next, expanding the fermion in terms of the KK modes, 
\bea
% \bar\Psi(x^\mu,y) &=& \frac1{\sqrt{2\pi R}} \sum_{n=-\infty}^\infty 
%                        \bar\Psi^{(-n)}(x^\mu) 
%                        \exp \left(i n\frac{y}{R} \right), \\
 \Psi(x^\mu,y) = \frac1{\sqrt{2\pi R}} \sum_{n=-\infty}^\infty 
                    \Psi^{(n)}(x^\mu) \exp \left(i n\frac{y}{R} \right),
\eea
and integrating out $y$ coordinate, the fermion part is written 
as\footnote{In the case of odd $D$, $\Psi^{(n)}(x^\mu)$ in (\ref{fermion}) 
represents two $D$ dimensional spinors, simultaneously.} 
\bea
 S_m = \int d^D x \sum_{m,n}
            {\bar\Psi^{(m)}}
            \Ls i \delta_{nm}\Ls \sla\del + i M_{n+\alpha} \Rs  
                + \sum_{l}\delta_{m\ l+n} 
                 \Ls g_D \sla\! A_\mu^{(l)} + g_D A_y^{(l)}\Rs
            \Rs\Psi^{(n)}
            \label{fermion}
\eea
where the $D$ dimensional gauge coupling constant 
$g_D$ is defined as $g_D=g/\sqrt{2\pi R}$. 
The zero-mode fermion is known to have a mass $M_\alpha = \alpha/R$. 
From (\ref{gauge}) and (\ref{fermion}), Feynman rules we need can be read off. 
%%%%%%%%%%%%%%%%%
\begin{figure}[h]
 \begin{center}
  \includegraphics[width=3.0cm]{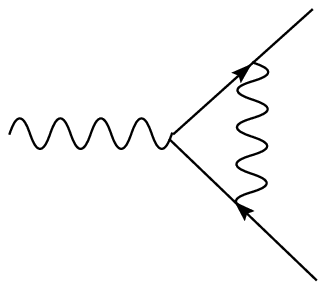}
  \hspace*{30mm}
  \includegraphics[width=3.0cm]{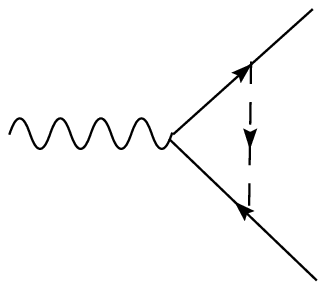}
  \put(-280,40){{\large $\gamma_\mu$}}
  \put(-175,75){{\large $\Psi$}}
  \put(-185,35){{\large $A_\mu^{(n)}$}}
  \put(-175,-05){{\large $\Psi$}}
  \put(-100,40){{\large $\gamma_\mu$}}
  \put(00,75){{\large $\Psi$}}
  \put(-10,35){{\large $A_5^{(n)}$}}
  \put(00,-05){{\large $\Psi$}}
  \put(-225,-10){$(A)$}
  \put(-55,-10){$(B)$}
 \end{center}
% \vspace*{-0.5cm}
\caption{The diagrams contributing to the anomalous magnetic moment 
in the $(D+1)$ dimensional QED on $M^D \times S^1$. 
The diagram (A)((B)) is the photon (Higgs) KK mode exchange diagram, 
respectively.} 
\label{mmm}
\end{figure}
%%%%%%%%%%%%%%%%%%%%%%

Now, we are ready to calculate the anomalous magnetic moment 
of the zero-mode fermion. 
The diagrams we should calculate are shown in Fig. \ref{mmm} 
and are calculated as 
\bea
(A) 
&=& \int \frac{d^Dk}{(2\pi)^Di}\sum_{n=-\infty}^\infty g_D^3
\frac{\gamma_\nu(p\!\!\!/' + k\!\!\!/ + M_{n+\alpha}) \gamma_\mu 
(p\!\!\!/ + k\!\!\!/ + M_{n+\alpha}) \gamma^\nu}
{[(p'+k)^2 - M^2_{n+\alpha}][(p+k)^2 - M^2_{n+\alpha}](k^2 - M_n^2)}, 
\label{A} \\
(B) 
&=& - \int \frac{d^Dk}{(2\pi)^Di}\sum_{n=-\infty}^\infty g_D^3
\frac{(p\!\!\!/' + k\!\!\!/ + M_{n+\alpha}) \gamma_\mu 
(p\!\!\!/ + k\!\!\!/ + M_{n+\alpha})}
{[(p'+k)^2 - M^2_{n+\alpha}][(p+k)^2 - M^2_{n+\alpha}](k^2 - M_n^2)}
\label{B}
\eea
where $p_\mu(p_\mu')$ is the external momentum of $\Psi(\bar{\Psi})$. 
Noting that the numerators can be cast into the forms consisting of 
$\gamma_\mu, p_\mu + p_\mu'$ by use of the on-shell condition of $\Psi$ 
and that the only $(p_\mu + p_\mu')$ term contributes to 
the anomalous magnetic moment, 
we obtain the contribution of each diagram to the anomalous magnetic moment 
$a \equiv (g-2)/2$, 
\bea
a(A) &\equiv& -\frac{2M_\alpha}{g_D} \times (A) \nonumber \\
&=& -4 g_D^2 M_\alpha 
\int \frac{d^Dk}{(2\pi)^Di}\sum_{n=-\infty}^\infty 
\int_0^1 dx \int_0^{1-x} dy
\times \nonumber \\
&& \frac{[2-(D-2)(x+y)](x+y)M_\alpha + [4-D(x+y)]M_n}
{[k^2 -(M_n + (x+y)M_\alpha)^2]^3}, 
\label{a} \\
a(B) &\equiv& -\frac{2M_\alpha}{g_D} \times (B) \nonumber \\
&=& -4g_D^2 M_\alpha 
\int \frac{d^Dk}{(2\pi)^Di}\sum_{n=-\infty}^\infty 
\int_0^1 dx \int_0^{1-x} dy 
\frac{(x+y)(M_n + (2 - x - y)M_\alpha)}
{[k^2 - (M_n + (x+y)M_\alpha)^2]^3}
\label{b}
\eea
where $x,y$ are Feynman parameters and $g_D/(2M_\alpha)$ is a Bohr magneton. 
We note that the vertex under consideration are sandwiched 
between the wave functions of zero-mode fermion $\bar{\psi}^{(0)}(p')$ 
and $\psi^{(0)}(p)$. 
If $p\!\!\!/(p\!\!\!/')$ operates to $\psi^{(0)}(p)(\bar{\psi}^{(0)}(p'))$ 
from the left (right), 
it can be replaced by $- g \langle A_y \rangle = \alpha/R = M_\alpha$ 
because of the equations of motion. 
To obtain the final result, we make use of the symmetry under 
$x \leftrightarrow y$. 
%and symmetrized. 

As a check, if we consider the case with $n=0, D=4$, 
we can confirm that $a(A)$ and $a(B)$ coincide with the results 
of the anomalous magnetic moment due to the photon and Higgs 
exchange diagrams in the Standard Model \cite{FLS} 
provided the Higgs mass and Yukawa coupling are set to zero and $g_D$, 
respectively.

Let us first calculate $a(A)$ and $a(B)$ by taking the mode sum 
before the momentum integral. 
We use the following formula. 
\bea
S_1(x,\alpha) &\equiv& \sum_{n=-\infty}^\infty 
\frac{1}{[x^2+(\alpha + 2n \pi)^2]^3} \nonumber \\
&=& \frac{1}{16x} 
\left[
\frac{x^2 \sinh x -3(x \cosh x - \sinh x)}{x^4(\cosh x - \cos \alpha)} 
-\frac{3\sinh x(x \cosh x - \sinh x)}{x^3(\cosh x - \cos \alpha)^2} 
\right. \nonumber \\
&& \left. + \frac{2\sinh^3 x}{x^2(\cosh x - \cos \alpha)^3}
\right] \\
&\to& \frac{3}{16x^5}~(x \to \infty), 
\label{s1div} \\
S_2(x,\alpha) &\equiv& 
\sum_{n=-\infty}^\infty \frac{2n \pi}{[x^2 + (\alpha + 2n \pi)^2]^3} 
\nonumber \\
&=& - \frac{(x\cosh x - \sinh x)\sin \alpha}{16x^3(\cosh x - \cos \alpha)^2} 
+ \frac{\sinh^2 x \sin a}{8x^2(\cosh x - \cos \alpha)^3} 
- \alpha S_1(x,\alpha)
\label{s2} \\
&\to& -\alpha S_1(\infty, \alpha)~(x \to \infty). 
\label{s2div}
\eea
From these observations, 
it is useful to separate these functions into 
the possibly divergent part $S^{({\rm div})}$ 
and the super-convergent part $S^{({\rm sc})}$ as 
\bea
S_1^{({\rm div})}(x,\alpha) &=& \frac{3}{16x^5}, \\
S_1^{({\rm sc})}(x,\alpha) &=& 
\frac{1}{16x}
\left[
\frac{3}{x^4} \left( \frac{\sinh x}{(\cosh x - \cos \alpha)} -1 \right)
-\frac{3\cosh x}{x^3(\cosh x - \cos \alpha)} \right. \nonumber \\
&& \left. 
- \frac{2\sinh x}{x^2(\cosh x -\cos \alpha)} 
- \frac{3\sinh x \cos \alpha}{x^2(\cosh x - \cos \alpha)^2} \right. 
\nonumber \\
&& \left. 
+ \frac{3 \sinh^2 x}{x^3(\cosh x - \cos \alpha)^2} 
+ \frac{2\sinh^3 x}{x^2(\cosh x - \cos \alpha)^3} 
\right], \\
S_2^{({\rm div})}(x,\alpha) &=& -\alpha S_1^{({\rm div})}(x,\alpha), \\
S_2^{({\rm sc})}(x,\alpha) &=& 
- \frac{(x\cosh x - \sinh x)\sin \alpha}{16x^3(\cosh x - \cos \alpha)^2} 
+ \frac{\sinh^2 x \sin \alpha}{8x^2(\cosh x - \cos \alpha)^3} 
-a S_1^{({\rm sc})}(x,\alpha). 
\eea
From these quantities, the possibly divergent part of (\ref{a}) 
and (\ref{b}) can be read as
\bea
a(A)_{{\rm div}} &=& 8g_D^2 M_\alpha 
\int \frac{d^D k_E}{(2\pi)^D} \int_0^1 dt 
(2\pi R)^5(2\pi \alpha) t^2(-1+t) 
S_1^{({\rm div})} \left(2\pi k_E R, 2\pi t \alpha \right), \\
a(B)_{{\rm div}} &=& 8 g_D^2 M_\alpha 
\int \frac{d^Dk_E}{(2 \pi)^D} \int_0^1 dt 
(2\pi R)^5(2\pi \alpha) t^2(1-t)
S_1^{({\rm div})}
\left(2\pi k_E R, 2\pi t \alpha \right).
\eea
where $x+y \equiv t$ and $k_E$ is an Euclidean momentum. 
Clearly, we immediately see $a(A)_{{\rm div}} + a(B)_{{\rm div}} =0$. 
We found that the possibly divergent part is exactly 
canceled irrespectively of the dimensionality $D$ 
and a Feynman parameter $t$, as expected.

It is straightforward to obtain the explicit expression 
for the anomalous magnetic moment as the sum of 
remaining super-convergent parts as
\bea
a(A)_{{\rm sc}} + a(B)_{{\rm sc}} 
&=&
4 g_D^2 M_\alpha \int \frac{d^Dk_E}{(2\pi)^D} \int_0^1 dt 
(2\pi R)^5 t(4-(D-1)t) \sin(2\pi t \alpha) \nonumber \\
&&\times \left[
-\frac{1}{16(2\pi R k_E)^2[\cosh(2\pi R k_E) - \cos(2\pi t \alpha)]} 
\right. \nonumber \\
&& \left. 
- \frac{((2\pi R k_E)\cos (2\pi t \alpha ) - \sinh (2\pi R k_E))}
{16(2\pi R k_E)^3[\cosh (2\pi R k_E) - \cos (2\pi t \alpha)]^2} 
\right. \nonumber \\
&& \left. 
+ \frac{\sinh^2 (2\pi R k_E)}
{8(2\pi R k_E)^2[\cosh (2\pi R k_E) - \cos (2\pi t \alpha)]^3} 
\right]. 
\label{sc}
\eea
Let us perform the momentum integral in (\ref{sc}). 
In order to do this, we need some formula in which 
the integrand is rewritten by the sum as
\bea
\frac{1}{x^2(\cosh x -\cos \alpha)} 
&=& 
%\frac{2}{x^2}\frac{e^{-x}}{e^{-2x}-2e^{-x}\cos \alpha +1} = 
\frac{2}{x^2}\sum_{n=1}^\infty \frac{\sin n \alpha}{\sin \alpha} 
e^{-nx}, \\
\frac{1}{x^2(\cosh x - \cos \alpha)^2} 
%&=& - \frac{1}{x^2\sin \alpha}\frac{\partial}{\partial \alpha}
%\left(\frac{1}{\cosh x - \cos \alpha} \right) \nonumber \\
&=& -\frac{2}{x^2\sin \alpha}\sum_{n=1}^\infty 
\frac{n \cos n \alpha \sin \alpha -\sin n \alpha \cos \alpha}{\sin^2 \alpha}
e^{-nx}, \\
\frac{\sinh x}{x^3(\cosh x - \cos \alpha)^2} 
&=& 
\frac{2}{x^3} \sum_{n=1}^\infty n\frac{\sin n \alpha}{\sin \alpha}e^{-nx}, \\
\frac{\sinh^2 x}{x^2(\cosh x - \cos \alpha)^3} 
&=& 
\frac{1}{x^2} \sum_{n=1}^\infty 
\left[
\frac{\sin n \alpha}{\sin^3 \alpha} 
- \frac{n \cos n \alpha \cos \alpha}{\sin^2 \alpha} 
+ n^2 \frac{\sin n \alpha}{\sin \alpha}
\right]e^{-nx}. 
\eea
Combining these results, we obtain the final expression 
for the anomalous magnetic moment, 
\bea
a(A)_{{\rm sc}} + a(B)_{{\rm sc}} 
= 
\frac{g_D^2 \Gamma(D-1) M_\alpha}
{(2\pi R)^{D-5}(4\pi)^{D/2}\Gamma(\frac{D}{2})(D-3)} 
\int_0^1 dt t(4-(D-1)t) \sum_{n=1}^\infty \frac{\sin(2\pi n t \alpha)}{n^{D-4}} 
\label{scvalue} 
\eea
where 
\bea
\Gamma(z) = \int_0^\infty dt t^{z-1} e^{-t}
\eea
is used.

As a consistency check, let us evaluate the finite value 
of the anomalous magnetic moment 
by an alternative method, {\em i.e.} 
by doing the momentum integral before the mode sum 
invoking Poisson resummation, 
\bea
a(A) &=& 4g_D^2 M_\alpha \int_0^1 dt \int \frac{d^Dk}{(2\pi)^D} 
\sum_{n=-\infty}^\infty 
\left[
t^2(2-(D-2)t)M_\alpha + t(4-Dt)M_n
\right] \nonumber \\
&& \times \int_0^\infty ds \frac{s^2}{\Gamma(3)}e^{-[k^2 + (M_n + tM_a)^2]s} 
\nonumber \\
&=& 2g_D^2 M_\alpha \int_0^1 dt \int_0^\infty ds 
\frac{s^{2 - \frac{D}{2}}}{(4\pi)^{D/2}} \times \nonumber \\
&&\sum_{n=-\infty}^\infty 
\left[
t^2 (-2+2t)M_a R \sqrt{\frac{\pi}{s}} 
+t(4-Dt)R^2 \sqrt{\frac{\pi}{s^3}}(i\pi n)
\right] e^{-\frac{(\pi Rn)^2}{s} -2\pi i n t \alpha}
\label{Alternative}
\eea
where $t$ is a Feynman parameter. 
In the first line, we used the integral expression for the Gamma function 
\bea
\frac{1}{\Delta^s} = \int_0^\infty dt \frac{t^{s-1}}{\Gamma(s)}e^{-\Delta t}.
\eea
In the second line, the Gaussian integral for the momentum is performed and 
Poisson resummation formulas listed below are used, 
with the replacement $m \to n$ being done afterwards, 
\bea
\sum_{n=-\infty}^\infty e^{-(\frac{n+t \alpha}{R})^2 s} 
&=& \sum_{m=-\infty}^\infty R\sqrt{\frac{\pi}{s}} 
e^{-\frac{(\pi Rm)^2}{s}-2\pi im \alpha t}, \\
\sum_{n=-\infty}^\infty \left(\frac{n+t \alpha}{R} \right) 
e^{-(\frac{n+t \alpha}{R})^2 s} 
&=& \sum_{m=-\infty}^\infty R^2 \sqrt{\frac{\pi}{s^3}}(i\pi m) 
e^{-\frac{(\pi Rm)^2}{s}-2\pi im \alpha t}. 
\eea
For $n=0$ case in the second line of (\ref{Alternative})
(``zero-winding" sector \cite{CMS}), 
the integral concerning $s$ diverges at $s=0$, 
which gives the divergent part 
\bea
\tilde{a}(A)_{{\rm div}} 
= -\frac{g_D^2 \alpha M_\alpha \sqrt{\pi}}{3(4\pi)^{D/2}} 
\int_0^\infty ds s^{\frac{3-D}{2}}. 
\label{divA}
\eea
The remaining finite part ($n \ne 0$) is 
\bea
a(A)_{{\rm finite}} 
&=& \frac{2g_D^2 \sqrt{\pi}M_\alpha}{(4\pi)^{D/2}} \int_0^1 dt 
\Gamma \left(\frac{D-5}{2} \right) \times \nonumber \\
&& \sum_{n=1}^\infty
\left[
4t^2(-1+t)\alpha \frac{\cos(2\pi nt \alpha)}{(\pi Rn)^{D-5}} 
+\left(\frac{D-5}{2} \right) 
2t(4-Dt)R \frac{\sin(2\pi nt \alpha)}{(\pi R n)^{D-4}}
\right]. 
\label{finiteA}
\eea
Similarly to the case $(A)$, the corresponding 
divergent and finite parts of $(B)$ 
are calculated to be 
\bea
\tilde{a}(B)_{{\rm div}} 
%&=& -g_D^3 \int_0^1 dt \int_0^\infty ds 
%\frac{s^2t^2}{(4\pi s)^{D/2}} 2(1-t)a\sqrt{\frac{\pi}{s}} 
&=& \frac{g_D^2 \alpha M_\alpha \sqrt{\pi}}{3(4\pi)^{D/2}} \int_0^\infty ds 
s^{\frac{3-D}{2}} = -\tilde{a}(A)_{{\rm div}}, \\
a(B)_{{\rm finite}} 
&=& g_D^2 \frac{2 M_\alpha \sqrt{\pi}}{(4\pi)^{D/2}} \int_0^1 dt 
\Gamma \left(\frac{D-5}{2} \right) \times \nonumber \\
&& \sum_{n=1}^\infty 
\left[
4t^2(1-t)\alpha \frac{\cos(2\pi nt \alpha)}{(\pi Rn)^{D-5}} 
+2t^2R \frac{\frac{D-5}{2}\sin(2\pi nt \alpha)}{(\pi Rn)^{D-4}}
\right] 
\eea
where we can confirm that the divergences from $(A)$ and $(B)$ 
are exactly canceled, 
namely $\tilde{a}(A)_{{\rm div}} + \tilde{a}(B)_{{\rm div}} = 0$. 
Thus, we obtain the final result for the finite anomalous magnetic moment 
\bea
a = a(A)_{{\rm finite}} + a(B)_{{\rm finite}} 
= 
\frac{4g_D^2 \Gamma(\frac{D-3}{2})M_\alpha}
{(4\pi)^{D/2}\sqrt{\pi}(\pi R)^{D-5}}
\int_0^1 dt t(4-(D-1)t) 
\sum_{n=1}^\infty \frac{\sin(2\pi nt \alpha)}{n^{D-4}}. 
\label{finiteAB} 
\eea
One can easily check $a(A)_{{\rm finite}} + a(B)_{{\rm finite}} =
a(A)_{{\rm sc}} + a(B)_{{\rm sc}}$.

To see how the contribution of non-zero KK modes behaves, 
it will be useful to expand the final result (\ref{finiteAB})
(or (\ref{scvalue})) in terms of the ratio of the fermion mass to 
the compactification scale $M_\alpha/(R^{-1}) = \alpha$ 
under a plausible assumption $\alpha \ll 1$. 
We note that the sum over $n$ in (\ref{finiteAB}) is exactly given 
for $D=4$ as 
\bea
\sum_{n=1}^\infty \sin(2\pi n t \alpha) 
= \frac{1}{2}\cot(\pi t \alpha) 
\simeq \frac{1}{2\pi t \alpha} -\frac{\pi t \alpha}{6}. 
\eea
Thus, for $D=4$ (five dimensional spacetime), 
the anomalous magnetic moment is given as 
\bea
a(D=4) \simeq \frac{5g_4^2}{16\pi^2} - \frac{7g_4^2}{288}\alpha^2. 
\eea
The first term corresponds to the zero mode contribution, 
though it does not completely agree with the result 
in four dimensional QED, since in our case the contribution of 
the Higgs exchange diagram is comparable to that of photon exchange diagram. 
The second term of ${\cal O}(\alpha^2)= {\cal O}(M_\alpha^2/R^{-2})$ is 
the contribution of non-zero KK modes, which is suppressed 
by the inverse power of $1/R$: the ``decoupling" of massive KK modes.

In the case with spacetime dimension higher than five ($D > 4$), 
the separation of the contribution of non-zero KK modes is not straightforward. 
In fact, the leading order term in the power series expansion of $\alpha$ 
does not correspond to the four dimensional result. 
For instance, for $D=5$ the leading term behaves as $\sim g_5^2 M_\alpha$, 
since 
\bea
\sum_{n=1}^\infty \frac{\sin(2\pi n t \alpha)}{n} 
= \frac{\pi}{2} (1 - 2 t \alpha). 
\eea
The origin of this problem is that we are assuming only one space dimension 
is compactified on $S^1$, even for six dimensional spacetime ($D=5$). 
To get a meaningful result, it will be necessary to work 
in a realistic situation where the extra space is, say, 
$T^2$ with KK modes $(n, n')$, and the four dimensional contribution can be 
extracted as the contribution of $(0, 0)$ sector. 
However, we leave this issue for a future publication, 
since our main purpose in this article is to show the finiteness of 
the anomalous magnetic moment, not its precise value.

In summary, we have shown that 
the anomalous magnetic moment is UV finite and predictive 
in a $(D+1)$ dimensional QED gauge-Higgs unification model compactified on $S^1$. 
This result is naturally expected 
because the local gauge invariant operator relevant 
for the anomalous magnetic moment is forbidden by the higher dimensional 
gauge invariance, the Lorentz invariance and the on-shell condition. 
The finite value was derived by two independent methods, 
{i.e.},  taking first the KK mode sum 
before the momentum integral and vice versa. 

Some comments are in order. 
We have considered the case where only one of the spatial 
dimension is compactified. 
We note that the argument for UV finiteness discussed here is unchanged 
even if we consider a more realistic compactification 
though the finite value might be changed. 
This is because the compactification affects 
only the IR physics not the UV one.

The anomalous magnetic moment of the muon has been also calculated 
in the universal extra dimension (UED) scenario \cite{UEDg-2}, 
where it is finite in five dimensional spacetime, 
but divergent for more than five dimensions. 
This is the natural result from the argument of power counting. 
Thus, we can see that the predictions for the anomalous magnetic moment 
from the gauge-Higgs unification 
and UED scenarios are quite distinct.

The final comment is a concern about the electric dipole moment 
of fermion in the gauge-Higgs unification. 
From the viewpoint of operator analysis, we notice that 
the only difference of the gauge invariant local operator 
relevant for the electric dipole moment from that for the magnetic moment 
is a $\Gamma_y$ insertion. 
This immediately suggests that the gauge invariant operator 
for the electric dipole moment is also forbidden 
due to the higher dimensional gauge invariance and the on-shell condition, 
as discussed in this letter. 
Therefore, we can expect that the electric dipole moment 
in the gauge-Higgs unification also becomes finite in any spacetime dimensions 
if it ever exists.

%%%%%%%%%%%%%%%%%%%%%%%%%%%%
\subsection*{Acknowledgment}
%%%%%%%%%%%%%%%%%%%%%%%%%%%%
The work of the authors was supported 
in part by the Grant-in-Aid for Scientific Research 
of the Ministry of Education, Science and Culture, No.18204024.


\begin{thebibliography}{99}

%%%%
\bibitem{MF} 
  N.~S.~Manton,
  %``A New Six-Dimensional Approach To The Weinberg-Salam Model,''
  Nucl.\ Phys.\ B {\bf 158}, 141 (1979); 
%\bibitem{Fairlie}
  D.~B.~Fairlie,
  %``Higgs' Fields And The Determination Of The Weinberg Angle,''
  Phys.\ Lett.\ B {\bf 82}, 97 (1979); 
  %``Two Consistent Calculations Of The Weinberg Angle,''
  J.\ Phys.\ G {\bf 5}, L55 (1979). 

%%%%%%%%%%%%%%%%%%%%%%%%%%%%%%%
\bibitem{Hosotani}  
  Y.~Hosotani,
%    %``Dynamical Mass Generation By Compact Extra Dimensions,''
  Phys.\ Lett.\ B {\bf 126}, 309 (1983); 
  %``Dynamical Gauge Symmetry Breaking As The Casimir Effect,''
  Phys.\ Lett.\ B {\bf 129}, 193 (1983); 
  %``DYNAMICS OF NONINTEGRABLE PHASES AND GAUGE SYMMETRY BREAKING,''
  Annals Phys.\  {\bf 190}, 233 (1989).

%%%%%%%%%%%%%%% 

%%%%%%%%%%%%%%%%%%%%%%%%%%%%%%%%%%%%%%%%%%%%%%%%%%%%%%%%%%%%%%% 
\bibitem{HIL}
  H.~Hatanaka, T.~Inami and C.~S.~Lim,
  %``The gauge hierarchy problem and higher dimensional gauge theories,''
  Mod.\ Phys.\ Lett.\ A {\bf 13}, 2601 (1998). 
%  [arXiv:hep-th/9805067].
%%%%%%%%%%%%%%%%%%%%%%%%%%%%%%%%%%%%%%%%%%%%%%%%%%%%%%%%%%%%%%
\bibitem{ABQ}
  I.~Antoniadis, K.~Benakli and M.~Quiros,
  %``Finite Higgs mass without supersymmetry,''
  New J.\ Phys.\  {\bf 3}, 20 (2001). 
%  [arXiv:hep-th/0108005].


\bibitem{GIQ}
  G.~von Gersdorff, N.~Irges and M.~Quiros,
  %``Bulk and brane radiative effects in gauge theories on orbifolds,''
  Nucl.\ Phys.\ B {\bf 635}, 127 (2002). 
%  [arXiv:hep-th/0204223].

\bibitem{HLM}
  K.~Hasegawa, C.~S.~Lim and N.~Maru,
  %``An attempt to solve the hierarchy problem based on gravity gauge Higgs
  %unification scenario,''
  Phys.\ Lett.\ B {\bf 604}, 133 (2004). 
%  [arXiv:hep-ph/0408028].

\bibitem{MY}
  N.~Maru and T.~Yamashita,
  %``Two-loop calculation of Higgs mass in gauge-Higgs unification: 
  %5D  massless QED compactified on S**1,''
  Nucl.\ Phys.\ B {\bf 754}, 127 (2006). 
%  [arXiv:hep-ph/0603237].

\bibitem{LMH}
  C.~S.~Lim, N.~Maru and K.~Hasegawa,
  %``Six dimensional gauge-Higgs unification with an extra space S**2 and the
  %hierarchy problem,''
  arXiv:hep-th/0605180.

%%%%% 
\bibitem{Hosotani2}
  Y.~Hosotani,
  %``All-order finiteness of the Higgs boson mass in the dynamical gauge-Higgs
  %unification,''
  arXiv:hep-ph/0607064.



%%%%%%%%%%%%%%% 
\bibitem{LM}
  C.~S.~Lim and N.~Maru,
  %``Calculable One-Loop Contributions to S and T Parameters in the
  %Gauge-Higgs Unification,''
  Phys.\ Rev.\  D {\bf 75}, 115011 (2007). 
%  [arXiv:hep-ph/0703017]

\bibitem{GIQ}
  G.~von Gersdorff, N.~Irges and M.~Quiros,
  %``Bulk and brane radiative effects in gauge theories on orbifolds,''
  Nucl.\ Phys.\  B {\bf 635}, 127 (2002); 
  %[arXiv:hep-th/0204223].
  %``Finite mass corrections in orbifold gauge theories,''
  arXiv:hep-ph/0206029. 


\bibitem{FLS}
  K.~Fujikawa, B.~W.~Lee and A.~I.~Sanda,
  %``Generalized Renormalizable Gauge Formulation Of Spontaneously Broken Gauge
  %Theories,''
  Phys.\ Rev.\  D {\bf 6}, 2923 (1972).

%%%%% 
\bibitem{CMS}
  H.~C.~Cheng, K.~T.~Matchev and M.~Schmaltz,
  %``Radiative corrections to Kaluza-Klein masses,''
  Phys.\ Rev.\  D {\bf 66}, 036005 (2002). 
%  [arXiv:hep-ph/0204342].

\bibitem{UEDg-2}
  K.~Agashe, N.~G.~Deshpande and G.~H.~Wu,
  %``Can extra dimensions accessible to the SM explain the recent  measurement
  %of anomalous magnetic moment of the muon?,''
  Phys.\ Lett.\  B {\bf 511}, 85 (2001); 
  T.~Appelquist and B.~A.~Dobrescu,
  %``Universal extra dimensions and the muon magnetic moment,''
  Phys.\ Lett.\  B {\bf 516}, 85 (2001).


\end{thebibliography}
\end{document}